\documentclass[pra,twocolumn,nofootinbib,showpacs,preprintnumbers,amsmath,amssymb]{revtex4}
\usepackage{graphicx}
\usepackage{dcolumn}
\usepackage{bm}
\usepackage[usenames]{color}
\usepackage{color}
\def\>{\rangle}\def\<{\langle}
\def\togli#1{}
\def\comment#1{}
\def\labell#1{\label{#1}}

\begin{document}
\title{Exploring Quantum Correlations from Discord to Entanglement}
\author{E. Moreva$^{1,3}$\footnote{corresponding author: ekaterina.moreva@gmail.com}, M. Gramegna$^{1}$, M. A. Yurischev$^{2}$}
\affiliation{\vbox{$^1$ Istituto Nazionale di Ricerca Metrologica, strada delle Cacce 91, 10135 Torino, Italy} \vbox{$^2$ Institute of Problems of Chemical Physics of the Russian Academy of Sciences,
142432 Chernogolovka, Moscow Region, Russia}\vbox{$^1$ Universitа di Torino, Dip.Fisica, via P. Giuria 1, 10125 Torino, Italy}}

\begin{abstract}
Quantum correlations represent a fundamental tool for studies ranging from basic science to quantum technologies. Different non-classical correlations have been identified and studied, as entanglement and discord. In this Paper we explore experimentally the rich geometry of polarization Bell-diagonal states. By taking advantage of the statistical method of generation, the values of entanglement and discord along different trajectories in the space of the parameters of density matrix have been measured. The effects of sudden death of entanglement and complete "freeze" of discord were investigated in order to detect the domains with different domination of one type of quantum correlation against to other. A geometric interpretation for each considered phenomena is addressed. The observed good agreement between experiment and theory for all investigated trajectories ensures the reliability of this method.
\end{abstract}

\maketitle

\section{Introduction}
Quantum correlations are now recognised to be a tool of capital importance for fundamental physics studies (\emph{i.e.} in foundations of quantum mechanics \cite{prep}, and also for researches ranging from phase transitions \cite{pt, WANG} to cosmology \cite{cosm, EmTime})  and, even more significantly, an essential resource for emerging quantum technologies as quantum communication \cite{qcom,qcom2,chir}, quantum computing \cite{Nielsen}, quantum metrology \cite{Girolami} and quantum imaging \cite{qimag}.
Despite quantum correlations were explained at the beginning exclusively in terms of quantum entanglement \cite{AFOV08,HHHH09,sar,zyc,epl}, in the recent years the concept of quantum discord deserved prominence \cite{Z00,HV01,OZ02,V03,Z03} as a more general and fundamental type of correlation with respect to entanglement (see also reviews \cite{CMS11,MBCPV12}). Quantum discord gauges the quantumness of correlations, being revealed by means of local operations acting a disturbance on the state. It turned out that the discord can appear even when separable states are considered, cases for which entanglement vanishes completely. Furthermore, it was shown that "almost all quantum states have nonclassical correlations" \cite{FACCA10}. On the consequence, discord represents a complementary quantum resource exploitable in particular in all those contexts in which characterization of quantum correlations results of non-trivial processing, in particular dealing with multipartite systems, mixed states and noisy environment. The increasing interest in quantum correlation dynamics and in the different behaviours between quantum discord and entanglement led to a deeper investigation in this field.

On the theoretical side, the characteristic feature of quantum correlations as a function of parameters defining the quantum state is a piecewise-defined form. The domain of definition is divided into subdomains with distinguished boundaries, at each subdomain corresponding its own branch of the function. The origin of the piecewise structure ensues from the optimization procedure, included in the definition of the quantum entanglement and discord. During the evolution, the quantum system moves along a certain trajectory in the parameter space, crossing the boundaries between subdomains which lead to sudden change of the branches of the quantum correlation function. In physical terms, it is the analogue to the phase transitions that occur, for example, in liquids or solids.

It has been shown that entanglement can suddenly disappear at a finite time, an effect defined as ”entanglement sudden death” \cite{YE04,YE06}. Sudden change of discord for two qubit states evolving through channels presenting different types of local decoherence was described in \cite{MCSV09,par1,par2}. Effects of robustness of quantum discord to sudden death and freezing phenomena were demonstrated in papers \cite{WSFB09,MPM10,LC10,AFA13}. Several peculiar properties in the dynamics of quantum correlations have been demonstrated experimentally. Entanglement sudden death, predicted for two-qubit entangled states interacting with noisy environments \cite{YE06, YE09}, was simulated and proved experimentally in \cite{dav,xu,ad,AMM07,SMA08,XLG10}. Specifically, in \cite{XLG10} the entanglement revival phenomenon was observed, and the freezing effect of quantum discord for two-photon state in a phase-dumping channel was investigated in the experimental work \cite{xu}.

In the direction to define operational measures to evaluate the complex interplay between entanglement and discord, a particular relevance assumes the class of two-qubit states that are diagonal in the Bell basis \cite{LC10}, and for which the nonclassical measures can be calculated explicitly, allowing at the same time the determination of changes in their behaviour under decoherence \cite{bdsi,bdsir,xu}.

In this paper we address the theoretical and experimental determination both of discord and entanglement and their properties for the family of Bell-diagonal states of two polarization photons along different trajectories in the space of the parameters of density matrix, as discussed in \cite{LC10}. The geometry of Bell-diagonal states is a three-parameter set, including the subsets of separable and classical subsets, and thus can be depicted in three dimensions. Level surfaces of entanglement and nonclassical measures can be plotted directly on this geometry, resulting into a complete picture of the structure of entanglement and nonclassicality able to describe their respective changes under the action of decoherence.
The statistical method of preparation of the given Bell-diagonal states allows to significantly extend the variety of trajectories in comparison with all-optical experimental setups. We explore peculiar behaviours of quantum entanglement and discord like sudden death, phase transition, freezing phenomena, and detect domains with different domination of one type of quantum correlation against to other. For each trajectory of evolution we present also a geometrical representation of piecewise quantum correlations functions.

\section{Measures of quantum correlation and geometric representation}

The entanglement of formation $E$ for arbitrary two-qubit systems \cite{wootters} is defined as,

\begin{equation}
   \label{eq:EConc}
   E=H((1+\sqrt{1-C^2})/2),
\end{equation}

where $H(x)=-x\log_2x-(1-x)\log_2(1-x)$ is the binary entropy function and concurrence $C$ is given by Hill-Wootters formula \cite{HW97,W98}
\begin{equation}
   \label{eq:CHW}
   C=\max\{0,\,\sqrt{\lambda_1} - \sqrt{\lambda_2}
   - \sqrt{\lambda_3} - \sqrt{\lambda_4}\}.
\end{equation}
Here $\lambda_i$ are eigenvalues of the matrix $R=\rho(\sigma_y\otimes\sigma_y)\rho^*(\sigma_y\otimes\sigma_y)$ in decreasing order, and $\sigma_y$ is the Pauli matrix.
The quantum discord $D$  of a bipartite system is quantified as the difference between two definitions of the mutual information \cite{Vedral,Nielsen,mp}: the original quantum mutual information  $I(\rho)$ and an alternative version of the quantum mutual information, that quantifies only classical correlations $J$ (see Refs.\cite{olliver, Henderson, VV03} for further explanations):
\begin{equation}
   \label{eq:D}
   D = S(\rho_B) - S(\rho)
	 + \min_{\{\Pi_k\}}\sum_kq_kS({\rm Tr}_B(\Pi_k\rho\Pi_k^+)/q_k),
\end{equation}
where $\rho_B={\rm Tr}_A\rho$, $S(\ldots)$ is the von Neumann entropy of corresponding state and $q_k={\rm Tr}(\Pi_k\rho\Pi_k^+)$ is the probability to obtaining the outcome $k$.

In the present work we investigate in particular the family of Bell-diagonal states of two photons:
\begin{eqnarray}
    \rho=p_1|\Phi^{+}><\Phi^{+}|+p_2|\Phi^{-}><\Phi^{-}|+\nonumber\\+ p_3|\Psi^{+}><\Psi^{+}|+p_4|\Psi^{-}><\Psi^{-}|,
\labell{eq:Bell mix}
\end{eqnarray}
being $|\Phi^{\pm}>= (|00>\pm|11>)/\sqrt{2}$ and $|\Psi^{\pm}>= (|01>\pm|10>)/\sqrt{2}$ are the Bell basis.
These states are appealing to both theoretical and experimental studies since they present a relatively simple structure, and at the same time they are particularly suitable for illustrative examples. The family of states (\ref{eq:Bell mix}) belongs to the class of X states \cite{x}, and may be written in terms of the Bloch decomposition:
\begin{equation}
\rho= \tfrac{1}{4}\left(\mathbf{1}+\sum_{i=1}^3 c_{i}\sigma_{i}\otimes\sigma_{i}\right),
\label{eq:Bloch_representation}
\end{equation}
where $\sigma_{i}$ are the Pauli operators, and the coefficients $c_i$ are related with the eigenvalues $p_{i}$ (eq. \ref{eq:Bell mix}) of matrix (\ref{eq:Bloch_representation}) by: $p_{1}=(1+c_{1}-c_{2}+c_{3})/4$, $p_{2}=(1-c_{1}+c_{2}+c_{3})/4$, $p_{3}=(1+c_{1}+c_{2}-c_{3})/4$, $p_{4}=(1-c_{1}-c_{2}-c_{3})/4$. The diagonal entries of the correlation tensor $c_{1}$, $c_{2}$ and $c_{3}$ define a three-dimensional Cartesian coordinate system.
Due to the non-negativity definition of density operators, all $p_i\ge0$ and therefore the values of coefficients $c_i$ have to lie in a tetrahedron ${\cal T}$ with vertices $v_1=(1,-1,1)$, $v_2=(-1,1,1)$, $v_3=(1,1,-1)$, and $v_4=(-1,-1,-1)$. The centers of the tetrahedron faces are $o_1=(1/3,1/3,1/3)$, $o_2=(-1/3,-1/3,1/3)$, $o_3=(-1/3,1/3,-1/3)$ and $o_4=(1/3,-1/3,-1/3)$.
\begin{figure}
\begin {center}
\includegraphics[width=1\columnwidth]{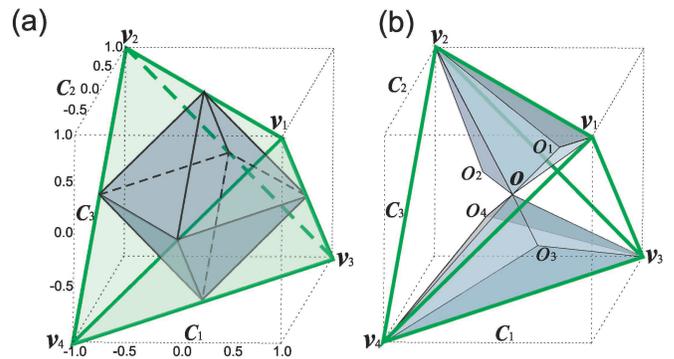}
\caption{(Color online) The total Bell diagonal states belong to the tetrahedron ${\cal T}$ with vertices $v_1$, $v_2$, $v_3$, and $v_4$. (a) The blue octahedron ${\cal O}$ inscribed in ${\cal T}$ is
a set of separable states. The remaining part $\cal T \backslash \cal O$ consists of four small tetrahedra $\tau_{i}$, where the entanglement is nonzero. (b) Two grey hexahedra show domain of definition of the branch $D_3$; domains $D_1$ and $D_2$ are similar but include the pairs of edges $\{v_1v_3,v_2v_4\}$ and $\{v_1v_4,v_2v_3\}$, correspondingly.}
\label{f:tetr_2}
\end {center}
\end{figure}
The concurrence for the quantum states (\ref{eq:Bell mix}, \ref{eq:Bloch_representation}) equals to:
\begin{equation}
   \label{eq:CBd}
   C=\max\{0,\,C_1, C_2, C_3, C_4\},
\end{equation}
where $C_{i}=2p_{i}-1$ $(i=1, ..., 4)$ \cite{BVSW96}.

The set of separable Bell-diagonal states lies inside the octahedron ${\cal O}$, inscribed into the tetrahedra $\tau_{i}$,  with $|c_1|+|c_2|+|c_3|\le1/2$ \cite{HH96}, while the four small tetrahedra $\tau_{i}$ outside of it belong to entangled Bell-diagonal states (Fig. \ref{f:tetr_2}(a)). Surfaces with constant entanglement belong to these small tetrahedra, and are parallel to the faces of octahedron ${\cal O}$. The entanglement has maximal values (equal to one) at the vertices $v_1$, $v_2$, $v_3$, $v_4$ of the tetrahedron ${\cal T}$, and monotonically decrease at moving off them. If the quantum system during its evolution enters into the domain ${\cal O}$, the entanglement suddenly  dies, on the contrary, when it leaves the octahedron, entanglement revives.

The quantum discord of Bell-diagonal states is described by an analytic formula of Luo \cite{Luo}:
\begin{equation}
   \label{eq:DLuo}
   D=\min\{D_1,D_2,D_3\},
\end{equation}
where
\begin{eqnarray}
   \label{eq:DiLuo}
   D_i=\sum_{k=1}^4p_k\log_2(4p_k)-\frac{1}{2}[(1-c_i)\log_2(1-c_i)+\nonumber\\+(1+c_i)\log_2(1+c_i)].
\end{eqnarray}

According to this formula, discord consists of three branches $D_1$, $D_2$, and $D_3$ (dubbed also as sub-functions, phases or fractures, in physical language), that correspond to three different situations: $|c_1|\ge\max\{|c_2|,|c_3|\}$, $|c_2|\ge\max\{|c_1|,|c_3|\}$, and $|c_3|\ge\max\{|c_1|,|c_2|\}$, splitting the tetrahedron ${\cal T}$ (as visible in Fig.\ref{f:tetr_2} (b)) by diagonal planes $|c_1|=|c_2|$, $|c_2|=|c_3|$, and $|c_3|=|c_1|$ \cite{Y15}.

Being discord a piecewise-defined function, likewise entanglement, when the system moves from one domain to another, on each of these boundaries it experiences a sudden change, which can be displayed as a fracture on its curve. The discord also reaches maximal values at the tetrahedron vertices and vanishes only on the Cartesian axes $Oc_1$, $Oc_2$, and $Oc_3$. This evidence is in agreement with a general conclusion \cite{FACCA10} that null-discord states are negligible in the whole Hilbert space.


\section{Experimental setup}

In order to generate the family of states (\ref{eq:Bell mix}) we used polarization two-qubit states \cite{prep,ave}, produced via spontaneous parametric down-conversion (set-up reported in Fig. 2). A set of two orthogonally oriented Type-I $\beta$-barium borate (BBO) crystals (1 mm), cut for collinear, non-degenerate frequency phase-matching around the central wavelength of $702$ nm, were pumped with 500 mW CW argon laser operating at $351$ nm. A Glan-Thompson prism set at vertical polarization (V) together with a half-wave plate $(\lambda_{p}/2)$ were used to rotate the polarization of the pump beam at $45^{\circ}$. Pairs of correlated photons with horizontal/vertical polarizations were generated in a coherent superposition

\begin{equation}
|\Phi>=\tfrac{1}{2}\left(|HH>+e^{i\varphi}|VV>\right).
\label{eq:Phi}
\end{equation}

An assebly of directable quartz plates (QP) controlled the relative phase shift $\phi$ between the horizontal and the vertical polarization components of the UV pump allowing the preparation of the two Bell states ($|\Phi^{+}>,|\Phi^{-}>$). To prepare the second pair of basis Bell states $|\Psi^{\pm}>$ an additional half-wave plate $(\lambda_{0}/2)$ was introduced in the transmitted arm after beamsplitter (BS). Set at an angle of $45^{\circ}$ the plate transformed the initial state $|\Phi^{\pm}>=\tfrac{1}{2}\left(|HH>\pm|VV>\right)$ into $|\Psi^{\pm}>=\tfrac{1}{2}\left(|HV>\pm|VH>\right)$.
In order to provide stability in the phase-matching conditions, the BBO crystals and QP were placed into a sealed box equipped with temperature control.
Dealing with the preparation of sets of states (\ref{eq:Bell mix}) with different coefficients $p_{1}$, $p_{2}$, $p_{3}$ and $p_{4}$ we used the following statistical method of preparation: the density matrix of the quantum state (\ref{eq:Bell mix}) from the mathematical point of view represents a sum of basis states $|\Phi^{\pm}>$, $|\Psi^{\pm}>$  with certain probabilities $p_{i}$. By changing the parameters of the setup and collecting statistics from the basis states with different acquisition times we can produce the desired state (\ref{eq:Bell mix}) with predetermined values of $p_{i}$. This statistical method \cite{shur} allows one to not be restricted by parameters of experimental setup and to prepare an arbitrary set of Bell-diagonal states with a high fidelity.
Finally, the prepared states were reconstructed using ququart state tomography \cite{kwiat2001,tomo,our}. In particular, in the preparation block of the set-up (Fig. 2) the photon pair constituting the biphoton-ququart was split into two spatial modes by using a beam splitter ($BS$). The projective measurements operated in each arm to perform the quantum state tomography were realized by means of a polarization filtering system consisting of a sequence of quarter- and half-wave plates, followed by a polarization prism, which transmits vertical polarization (V). The detection was operated by taking advantage of a silicon single-photon avalanche detectors (SPAD), with 30\% detection efficiency \cite{c1,c2}, on each arm, both connected to a coincidence electronic chain with a time window set at $1.5$ ns. By registering the coincidence rate for different projections, the reconstruction of the polarization density matrix of each incoming state has been performed. The obtained average fidelity for all states was greater than 98.8$\%$.

\begin{figure}[!ht]
\includegraphics[width=1\columnwidth]{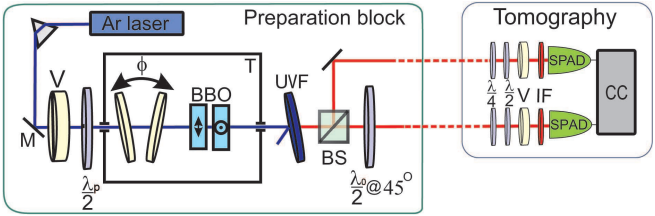}
\caption{(Color online) Experimental set-up. Preparation block: pairs of degenerate entangled
  photons are produced by pumping two orthogonally oriented type I BBO ($\beta-Ba B_2 O_4$) crystals. V is a Glan-Thompson prism with a vertical orientation, the half-wave plate
  $\lambda_{p}$/2 is oriented at an angle $\theta=45^{\circ}$, and UVF is an ultraviolet filter stopping the laser pump. Two 1 mm quartz plates, that can be rotated along the optical axis, introduce a phase shift $\phi$ between horizontally and vertically polarized photons.  The beam splitter (BS) is used to split the initial (collinear) biphoton field into distinct spatial modes. Additional half-wave plate $\lambda_{0}$/2 at $45^o$ in the transmitted arm is used to prepare the singlet Bell states $\Psi^{\pm}$. Tomography Block: Quarter-wave plates ($(\lambda/4)$), half-wave plates ($(\lambda/2)$) and Glan-Thompson prisms (V) are used in each mode to set the detecting bases for implementing quantum state tomography. Single-photon avalanche detectors (SPAD) equipped with interference filters (IF) are used to detect the photons.}
\label{f:setup}
\end{figure}

\begin{figure}[!ht]
\includegraphics[width=1\columnwidth]{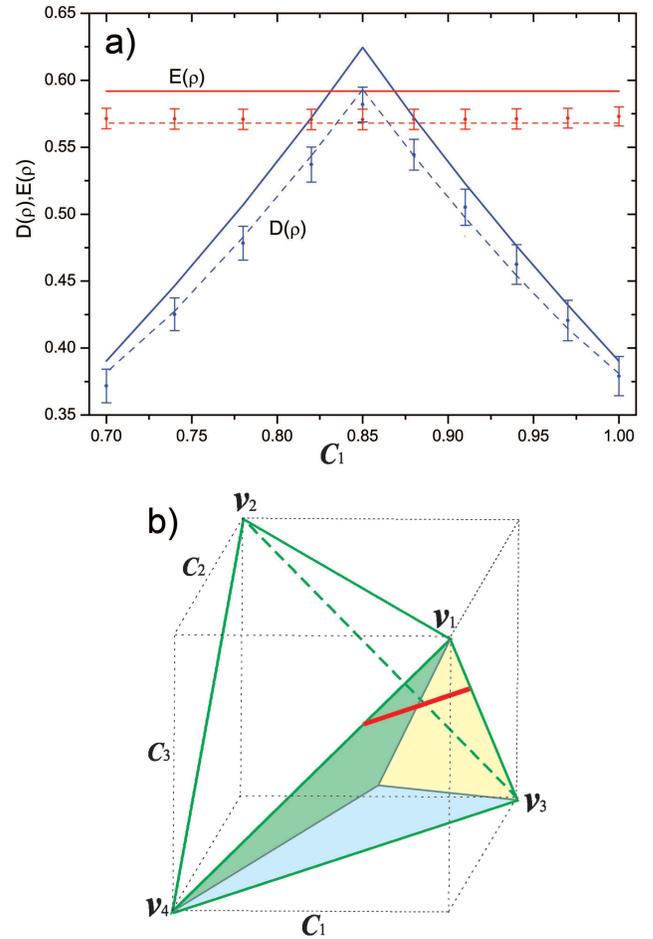}
\caption{(Color online) Graphs of $D(\rho)$ and $E(\rho)$ versus $c_{1}$ for the set of states defined by Eq.(\ref{eq:Bloch_representation}), with $c_2=c_1-1.7$, $c_3=0.7$. Fig. \ref{f:exp1}(a): While $E(\rho)$ is constant,  $D(\rho)$ can be lower or higher than $E(\rho)$. The deviation between the experimental data and the theoretical prediction is due to decoherence effects. In the depicted graph, solid lines correspond to a "pure" mix of basic Bell states, while dashed lines show theoretical states affected by decoherence process ($\nu=0.5\%$). Fig. \ref{f:exp1}(b) illustrates the tetrahedron of Bell diagonal states and the trajectory (red line) corresponding of the evolution. Three color domains (green, yellow and blue) correspond to discord with different phases $D_1$, $D_2$, $D_3$.}
\label{f:exp1}
\end{figure}

In order to reconstruct the value of both the discord and entanglement, a general approach requiring the density matrix only has been used. The theoretical behaviour of their dynamics were defined by analytical equations for Bell-diagonal states (\ref{eq:CBd}), (\ref{eq:DLuo}) and \ref{eq:DiLuo}). The experimental results were obtained from the reconstructed density matrices and general formulas (\ref{eq:EConc}), (\ref{eq:D}). The minimization over all possible projections was performed without any specific form of the quantum state. In fact, even if the fidelities of the reconstructed states are very high, the density matrices did not have exactly the Bell-diagonal form and contain, in general, $15$ independent real parameters.

\section{Results and discussion}

In order to obtain a deeper understanding in the interplay between the very rich dynamics respectively of discord and entanglement, and in the search of possible features to be exploited in technological applications, in this Paper we reported an experimental investigation of unique properties of the discord like sudden changes \cite{MCSV09}, freezing phenomenon \cite{LC10,MPM10} and robustness to sudden death \cite{WSFB09} with respect to sudden death of quantum entanglement \cite{YE09}, together with a full geometric interpretation of these phenomena. Special attention has been paid to the dynamics of quantum correlations against the uncontrollable effects due to decoherence or imperfection during the preparation, evolution or measurement stages. In particular, three sets were selected from the family of states (\ref{eq:Bell mix}), each of them presenting peculiar traits of discord or entanglement correlations to be analysed.

Firstly, let us consider the set of states lying on the red line belonging to the face $v_1v_3v_4$ of tetrahedron ${\cal T}$ and parallel to the edge $v_3v_4$ as  shown in  Fig. \ref{f:exp1}(b).
Assuming $c_3=0.7$, then the trajectory corresponds to $c_2=c_1-1.7$ where parameter $c_1\in[0.7,1]$. The theoretical behaviours of the discord and entanglement are shown in the Fig. \ref{f:exp1}(a) respectively by blue and red solid lines. The evolution of entanglement occurs along the constant line of the small tetrahedron $\tau_{1}$, meaning that during the entire evolution E($\rho$) is "freezed" and equal to $E=0.592$.

The behaviour of discord D($\rho$), on the other hand, moves from the domain ${\cal D}_2$ to domain ${\cal D}_1 $. It initially increases from $D=0.390$ to $D=0.624$, where in correspondence of $c_1=0.85$ a sharp bend is observed due to sudden transition, and then D($\rho$) decreases again to the original value $0.390$.

It is worth noting that over the two ranges $c_1\in[0.7,0.832]$ and $c_1\in[0.868,1]$ the value of E($\rho$) exceeds the discord -- a measure of the quantumness of correlations. This effect was predicted previously in Ref.~\cite{Luo} concerning with the Werner state (c1=c2=c3), but the maximal excess $E(\rho)$ over $D(\rho)$ was minimal and estimated about $2\% $. In the case considered in the present experiment the expected excess at $c_1=0.7$ (or 1) amounts to $52\%$.

In Fig. 3(a) the experimental points for $E(\rho)$ (red color) and $D(\rho)$ (blue color), together with their uncertainties, are reported with respect to theoretical curves. As it is visible, the data replicate the theoretical curves within uncertainties, but with a constant bias toward smaller values. This is due to a small decoherence process, related to imperfection at preparation stage. Indeed, the dashed lines, corresponding to theoretical calculations for the states affected by the decoherence process of $0.5\% $ are in excellent agreement with the reconstructed states. The value of decoherence was estimated by means of the visibility of two-photon interference.

\begin{figure}[!ht]
\includegraphics[width=1\columnwidth]{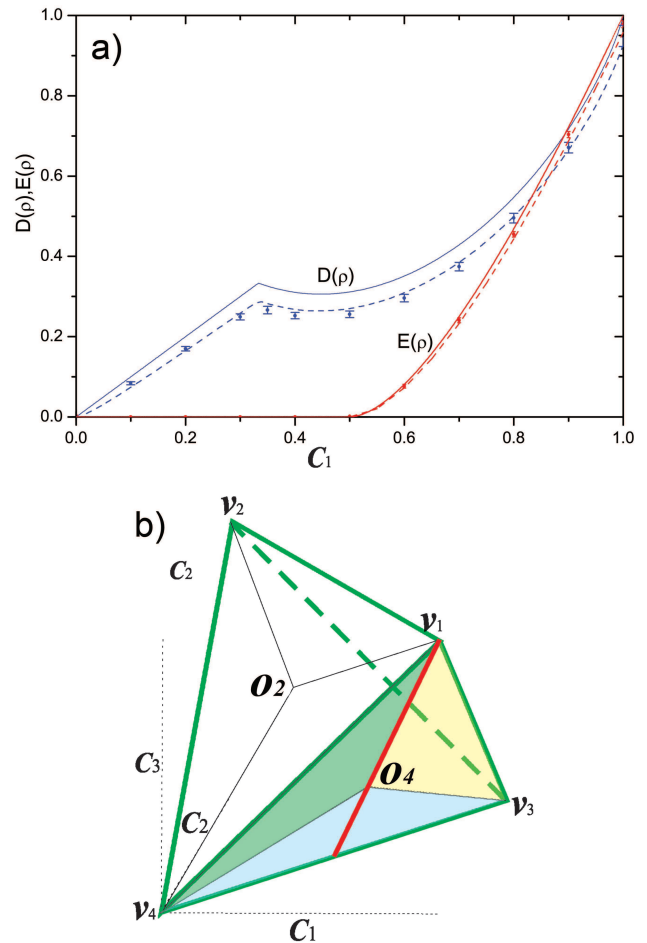}
\caption{(Color online) Graphs of $D(\rho)$ and $E(\rho)$ versus $c_{1}$ for the set of states defined by Eq.(\ref{eq:Bloch_representation}), with $c_2=-c_1$, $c_3=2c_1-1$ . Fig.(a): discord $D$ experiences a sudden changeat point $o_4$ , while entanglement decays monotonically up to sudden death at $c_1=0.5$. Here solid lines correspond to "pure" mix of basic Bell states, while dashed lines show theoretical states affected by decoherence process ($\nu=0.5\%$). Fig. ref{f:exp2}(b) shows tetrahedron of Bell states and the three color domains $D_1$, $D_2$, $D_3$ (green, yellow and blue) of discord with different phases. The trajectory (red line) runs along the border of the phases $D_1$ (green) and $D_2$ (yellow), and then enters on the region with discordant fracture $D_3$ (blue).}
\label{f:exp2}
\end{figure}

In Fig.~\ref{f:exp2}(a), the quantum discord and entanglement of formation are plotted as a function of $c_1$ when the three initial-state parameters satisfy $c_1=0:1$, $c_2=-c_1$ and $c_3=2c_1-1$. For this case, the evolution of Bell-mixed states occurs along the same face $v_1v_3v_4$, from the vertex $v_1$ to the middle of the edge $v_3v_4$ (Fig.\ref{f:exp2}(b)). On the site $v_1o_4$ trajectory runs along the border of the phases $D_1$ and $D_2$, and then enters into the region with discordant fracture $D_3$. At the point $o_4$ discord undergoes a sudden transition like a first-order phase transitions and then, with decreasing $c_1$, tends to zero according to the linear law $D=c_1$ (Fig.\ref{f:exp2}(a), blue line), meanwhile entanglement decays smoothly and has sudden death which happens at $c_1=0.5$.

\begin{figure}[!ht]
\includegraphics[width=1\columnwidth]{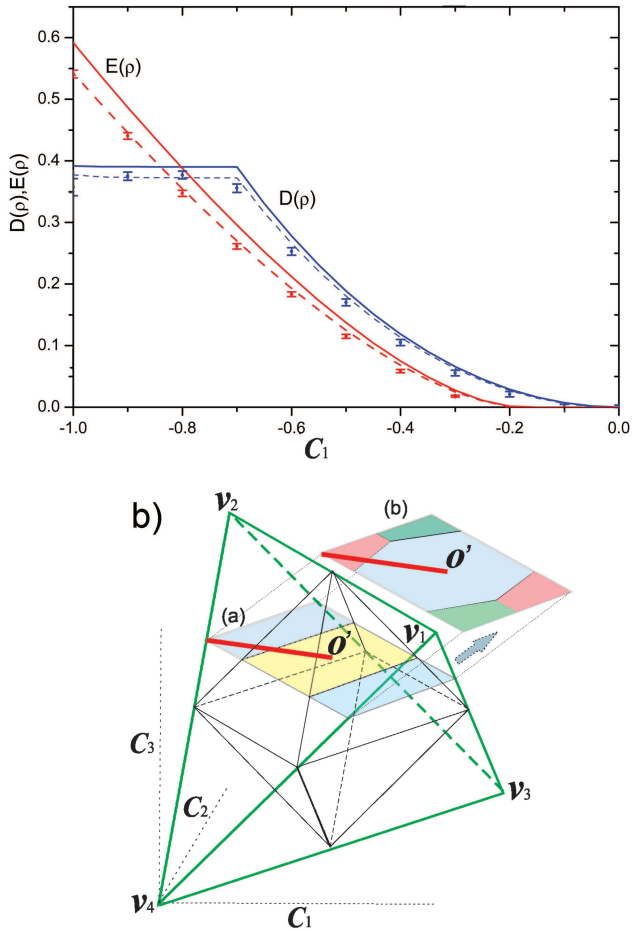}
\caption{(Color online) Fig.\ref{f:exp3}(a): Dynamics of entanglement $E(\rho)$ and discord $D(\rho)$ as a function of $c_1$ for the family states defined by Eq. (\ref{eq:Bloch_representation}), with parameters $c_3=0.7$, $c_2=-c_1c_3$. Here it is presented a freezing phenomenon of the discord. Dashed line shows a theoretical curve affected by decoherence process ($\nu=0.5\%$). Fig. \ref{f:exp3}(b) shows tetrahedron of Bell diagonal states and two section: section (a) shows regions of entangled (yellow) and separable (azure) states, while section (b) shows regions with different discord areas. The red line depicted over the sections represents the trajectory of the states under investigation.}
 \label{f:exp3}
\end{figure}

Another kind of correlation dynamics which simulates a phase-flip channel \cite{MCSV09,MPM10,LC10} is shown in Fig.~\ref{f:exp3}. The input state is given by equation (\ref{eq:Bloch_representation}) with parameters $c_1=-1:0$, $c_3=0.7$, $c_2=-c_1c_3$. The evolution of states corresponds to the thick red line in a perpendicular section to axis $c_3$, Fig.~\ref{f:exp3}(b). Regions of entangled (yellow) and separable (azure) states are shown in section (a), while section (b) shows regions with different discord areas and the trajectory of evolution from the corner to the center by the red line. The discord (blue line, Fig.~\ref{f:exp3}(a)) at such evolution remains constant, for a finite interval
of parametrized time $c_1\in[-1,-0.7]$, under decoherence, "freezed", and then decays monotonically, while the dynamics of the entanglement (red line), characterized by the entanglement of formation, smoothly relaxes and suddenly disappears at $c_1=-0.3/1.7\approx-0.176$. The blue and red circles in Fig.~\ref{f:exp3}(a) are the experimental results of $D(\rho)$ and $E(\rho)$ and correspond to theoretical predictions. The small deviations are explained by experimental imperfections and decoherence process.

\section{Summary}

In summary, in this work we have considered the geometrical approach of the quantum correlations in form of discord/entanglement and confronted it with the experimental observation. Selecting special trajectories of evolution with statistical preparation methodic of the Bell-diagonal states we demonstrated peculiar behaviours of entanglement (sudden death) and discord (phase transition, freezing phenomena). It was shown that entanglement can be frozen during the evolution, while the value of discord can be lower or higher. It follows that while the two quantities $E$ and $D$ measure
the same thing - the quantum correlation, differences in the actual functional dependencies mean that the relation between them is not monotone.
In our work we have been focused on straight trajectories, but this method allows one to reproduce the evolution of more complex, curved paths. The impact of these results, beyond their intrinsic interest, derives by demonstrating the experimental ability in controlling and manipulating these states, paving the way for the use of these resources in quantum technologies.

\section*{Acknowledgments}
\label{sec:ackn}
We acknowledge the support of  RFBR  under the Grant No.~15-07-07928. The results in this paper come from the project EMPIR 14IND05 MIQC2. This project has received funding from the EMPIR programme co-financed by the Participating States and from the European Union's Horizon 2020 research and innovation programme.

\end{document}